\documentclass{article}
\usepackage[T1]{fontenc}
\usepackage[utf8]{inputenc}
\usepackage[a4paper]{geometry}
\usepackage{mathrsfs}
\usepackage{amsmath}
\usepackage{hyperref}

\makeatletter
\@ifundefined{date}{}{\date{}}
\usepackage{authblk}
\usepackage{graphicx}

\title{Proof of the weak cosmic censorship conjecture for several extremal black  holes}

\author[1]{J\'{e}ssica Gon\c{c}alves}
\author[2]{Jos\'{e} Nat\'{a}rio\footnote{Corresponding author: \href{mailto:jnatar@math.tecnico.ulisboa.pt}{jnatar@math.tecnico.ulisboa.pt}}}

\affil[1]{School of Physics and Astronomy, University of Southampton, UK}

\affil[2]{CAMGSD, Mathematics Department, IST, University of Lisbon, Portugal}

\makeatother

\begin{document}
\maketitle 
\begin{abstract}
We show explicitly, for different types of extremal black holes, that
test fields satisfying the null energy condition at the event horizon
cannot violate the weak cosmic censorship conjecture. This is done
by checking, in each case, that the hypotheses for a general
theorem proved in a previous paper are satisfied. 
\end{abstract}

\section{Introduction}

In 1965, Penrose published the proof of the first singularity theorem
\cite{p28}, soon to be followed by increasingly more sophisticated results \cite{p29,p30}.
The singularities predicted by these theorems as a result of gravitational
collapse do not need, in principle, to be hidden inside a black hole
event horizon, that is, they could conceivably be naked singularities. This would
signal a major breakdown of general relativity, as the physics at
the singularity (and consequently in its causal future) cannot be
predicted by this theory.

In response to this problem, Penrose formulated, in 1969, the weak
cosmic censorship conjecture \cite{p31,p32}, hypothesizing that singularities
resulting from gravitational collapse must always be hidden inside
a black hole event horizon. This would prevent access to the singularity
by observers in the exterior of the black hole, meaning that the singularity
could not possibly influence the physics of the outside universe.

The weak cosmic censorship conjecture remains unproven. In an attempt
to violate it, Wald proposed a thought experiment to turn extremal
Kerr-Newman black holes into naked singularities by sending in spinning
and/or charged particles \cite{p33}. He discovered that particles with
large enough angular momentum or charge to overspin or overcharge
the black hole never went in, preserving the conjecture. The same
was found to be true for scalar and electromagnetic fields \cite{Semiz, Toth1, DuztasSemiz, Duztas1}, 
and, more generally, for any kind of test matter or fields \cite{p1}. A similar result also holds 
for near-extremal Kerr-Newman black holes, if backreaction
is taken into consideration \cite{p34}.

In this paper we explore the weak cosmic censorship conjecture further
by showing explicitly, using a general result in \cite{p1}, that
the conjecture holds for several other types of extremal black holes
which have been recently studied in the literature.

\section{Weak cosmic censorship conjecture}

\label{section2}

The proof in \cite{p1} that extremal Kerr-Newman or Kerr-Newman-AdS
black holes cannot be destroyed by interacting with test fields can
be adapted to other extremal black holes; this will be done in the
present paper in several examples. As already noted in \cite{p1},
the conditions that must be satisfied for the proof to go through
for a generic black hole are the following : 
\begin{enumerate}
\item \label{cond0} The Killing generator for the event horizon is of the
form 
\begin{equation}
Z=K+\sum_{i}\Omega_{i}Y_{i},
\end{equation}
where $K$ is an asymptotically timelike Killing vector field which
determines the physical mass of the black hole, $Y_{i}$ are angular
Killing vectors giving the angular momenta of the black hole, and
$\Omega_{i}$ are the thermodynamic angular velocities. 
\item The physical mass $M$, entropy $S$, angular momenta $J_{i}$ and
electric charge $Q$ can be related through a Smarr formula 
\begin{equation}
M=M(S,J_{i},Q),
\end{equation}
yielding a first law of black hole thermodynamics 
\begin{equation}
dM=TdS+\sum_{i}\Omega_{i}dJ_{i}+\Phi dQ,
\end{equation}
where $T$ is the black hole temperature and $\Phi$ is the event
horizon's electric potential. 
\item Extremal black holes (that is, black holes with $T=0$) are given
by $M=M_{ext}(J_{i},Q)$, and subextremal black holes by $M>M_{ext}(J_{i},Q)$. 
\item The test fields satisfy the null energy condition at the event horizon,
and appropriate boundary conditions at infinity (guaranteeing zero flux of conserved charges 
at infinity for suitably defined spacelike hypersurfaces). 
\item \label{cond} The interactions with the test fields preserve the black
hole type, that is, after the interaction the spacetime settles to
a stationary solution characterized by the same parameters $(M,J_{i},Q)$. 
\end{enumerate}
Condition \ref{cond} was not explicitly stated in \cite{p1}, but
will be important here, since we will be working with black holes
for which (in general) there are no uniqueness theorems. As a simple
example, when working with Reissner-Nordstr\"{o}m black holes only one
must consider interactions that preserve spherical symmetry (e.g.\ interactions
with spherically symmetric charged fields), as otherwise one may end
up with a Kerr-Newman spacetime. Even when working with Kerr-Newman
black holes, one should be aware that in general the black hole resulting
from the interaction with the test field is a Kerr-Newman spacetime
with a different symmetry axis, that is, that the black hole's angular
momentum may rotate; this is usually not discussed because only the
component of the absorbed angular momentum along the initial symmetry
axis is relevant. Indeed, if $(0,0,J)$ is the black hole's initial ADM
angular momentum and $(j_{1},j_{2},j_{3})$ is the absorbed
ADM angular momentum then, to first order, 
\begin{equation}
\left[(0,0,J)+(j_{1},j_{2},j_{3})\right]^{2}=J^{2}+2Jj_{3}=(J+j_{3})^{2},
\end{equation}
that is, the final spacetime has angular momentum $J+j_3$ along its 
new symmetry axis.

If conditions \ref{cond0} to \ref{cond} are satisfied then the test
fields cannot destroy the black hole, in the sense that if the black
hole absorbs energy $\Delta M$, angular momenta $\Delta J_{i}$ and
electric charge $\Delta Q$ by interacting with the test fields, then
the metric corresponding to the physical quantities $(M+\Delta M,J_{i}+\Delta J_{i},Q+\Delta Q)$
represents a subextremal (or at worse an extremal) black hole, rather
than a naked singularity. We now proceed to check that these conditions
hold for several extremal black holes in the literature which have
been recently studied in connection with the weak cosmic censorship
conjecture.

\section{BTZ black hole }

The BTZ metric is given by \cite{p2,p3,p4,p7} 
\begin{equation}
ds^{2}=-fdt^{2}+\frac{1}{f}dr^{2}+r^{2}\left(d\phi-\frac{J}{2r^{2}}dt\right)^{2},
\end{equation}
where 
\begin{equation}
f(r)=-M+\frac{r^{2}}{l^{2}}+\frac{J^{2}}{4r^{2}}
\end{equation}
and 
\begin{equation}
\Lambda=-\frac{1}{l^{2}}.
\end{equation}
Here $\Lambda$ is the cosmological constant, $M$ and $J$ are the
mass and angular momentum, respectively, and $r$ is the radial coordinate.
There are two horizons, which coincide when the black hole becomes
extremal (the inner horizon radius $r_{-}$ becomes equal to the outer
event horizon radius $r_{+}$). Since the metric is axisymmetric and
stationary, we have two Killing vectors, $X=\frac{\partial}{\partial t}$
and $Y=\frac{\partial}{\partial\phi}$, and thus the most general
Killing vector will be a linear combination of $X$ and $Y$.

The entropy $S$, angular velocity $\Omega$, Hawking temperature
$T$ and mass $M$ are given by \cite{p2,p3} 
\begin{equation}
S=4\pi r_{+},\hspace{1cm}\Omega=\frac{J}{2r_{+}^{2}},\label{BTZ1}
\end{equation}
\begin{equation}
T=\frac{r_{+}^{2}}{2\pi l^{2}}-\frac{J^{2}}{8\pi r_{+}^{3}},\hspace{1cm}M=\frac{r_{+}^{2}}{l^{2}}+\frac{J^{2}}{4r_{+}^{2}},\label{BTZ2}
\end{equation}
and satisfy the first law of black hole thermodynamics 
\begin{equation}
dM=TdS+\Omega dJ.
\end{equation}
The angular velocity of the black hole horizon is given by \cite{p5,p6}
\begin{equation}
\Omega_{H}=-\frac{g_{t\phi}}{g_{\phi\phi}}_{|_{r=r_{+}}}=\frac{J}{2r_{+}^{2}}\qquad\Rightarrow\qquad\Omega_{H}=\Omega,
\end{equation}
that is, the angular velocity of the event horizon is equal to the
thermodynamic angular velocity. Moreover, the angular velocity of
observers at infinity, with respect to the Killing vector $X$, is
given by 
\begin{equation}
\Omega_{\infty}=-\frac{g_{t\phi}}{g_{\phi\phi}}_{|_{r\rightarrow\infty}}=0,
\end{equation}
meaning that the asymptotically timelike Killing vector field which
determines the physical mass of the black hole is $K=X$. Therefore
the event horizon $\mathscr{H}^{+}$ Killing generator is given by
\begin{equation}
Z=X+\Omega_{H}Y=K+\Omega Y,
\end{equation}
and so the first condition of the theorem is satisfied

From \eqref{BTZ1} and \eqref{BTZ2} it is clear that the mass $M$
of the BTZ black hole can be entirely determined by the entropy $S$
and angular momentum $J$ of the event horizon through a Smarr formula
of the form 
\begin{equation}
M=M(S,J),
\end{equation}
meaning that the second condition of the theorem is also satisfied.

For the black hole to be extremal we must have $T=0$, which is equivalent
to 
\begin{equation}
\frac{\partial M}{\partial S}(S,J)=0.
\end{equation}
Solving this equation for the entropy leads to the entropy, hence
to the mass, of an extremal black hole as a function of its angular
momentum, 
\begin{equation}
S=S_{ext}(J)\Rightarrow M_{ext}=M(S_{ext}(J),J),
\end{equation}
which satisfies the third condition of the theorem.

If we assume that the test fields satisfy the null energy condition and preserve
the black hole type, then the five conditions of the theorem are satisfied.
Therefore, the BTZ black hole cannot be overspun to create a naked
singularity, that is, the weak cosmic censorship conjecture holds
for BTZ black holes.

\section{Quintessence RN-AdS black hole}

\label{sectionquintess}

The metric for the spherically symmetric Reissner-Nordstr\"{o}m-AdS black
hole surrounded by quintessence dark energy is given by \cite{p10}
\begin{equation}
ds^{2}=-fdt^{2}+\frac{1}{f}dr^{2}+r^{2}(d\theta^{2}+\sin^{2}\theta d\phi^{2}),
\end{equation}
where 
\begin{equation}
f(r)=1-\frac{2M}{r}+\frac{Q^{2}}{r^{2}}+\frac{r^{2}}{l^{2}}-\frac{a}{r^{3\omega+1}}\label{fquintess}
\end{equation}
and 
\begin{equation}
\Lambda=-\frac{3}{l^{2}}.
\end{equation}
Here $M$ and $Q$ are the mass and electric charge of the black hole,
respectively, $r$ is the radial coordinate, $a$ is a normalization
factor related to the density of quintessence dark energy, $l$ is
the AdS radius, $\Lambda$ is the cosmological constant, and $-1<\omega<-\frac{1}{3}$
so that we have quintessence dark energy.

Since the metric is static and spherically symmetric, the black hole
is nonrotating and there is no angular momentum to consider; consequently,
the thermodynamic angular velocity can be set to zero. The timelike
Killing vector $X=\frac{\partial}{\partial t}$ determines the black
hole mass and is also the Killing generator of the event horizon $\mathscr{H}^{+}$,
whose angular velocity is then also zero. In other words, $X=K=Z$,
and so the first condition of the theorem is satisfied.

We also have \cite{p10,p11} 
\begin{equation}
\Phi=\frac{Q}{r_{+}},\hspace{1cm}S=\pi r_{+}^{2},\label{entrquintess}
\end{equation}
\begin{equation}
T=\frac{f'(r_{+})}{4\pi}=\frac{1}{4\pi}\bigg[\frac{2M}{r_{+}^{2}}-\frac{2Q^{2}}{r_{+}^{3}}+\frac{2r_{+}}{l^{2}}+\frac{(3\omega+1)a}{r_{+}^{3\omega+2}}\bigg],
\end{equation}
where $r_{+}$, $\Phi$, $S$ and $T$ are the radius, electric potential,
entropy and Hawking temperature of the event horizon. For fixed $\Lambda$
and $a$, we have the first law of thermodynamics \cite{p10,p11,p12}
\begin{equation}
dM=TdS+\Phi dQ.
\end{equation}
Moreover, from \eqref{fquintess} it is clear that 
\begin{equation}
f(r_{+})=0\Leftrightarrow M=M(r_{+},Q),
\end{equation}
and using \eqref{entrquintess} we obtain a Smarr relation in the
form 
\begin{equation}
M=M(S,Q).
\end{equation}
Hence the second condition of the theorem is satisfied.

For the black hole to be extremal we must have $T=0$, which is equivalent
to 
\begin{equation}
\frac{\partial M}{\partial S}(S,Q)=0.
\end{equation}
Solving this equation for the entropy leads to the entropy, hence
to the mass, of an extremal black hole as a function of its charge,
\begin{equation}
S=S_{ext}(Q)\Rightarrow M_{ext}=M(S_{ext}(Q),Q),
\end{equation}
which satisfies the third condition of the theorem.

Finally, assuming that the test fields satisfy the null energy condition
and preserve the black hole type, we have just shown that it is not
possible to violate the weak cosmic censorship with such test fields,
as all conditions in the theorem are met.

\section{Gauss-Bonnet-AdS black hole}

We now apply the theorem to the Gauss-Bonnet-AdS black hole
in $d$ dimensions. The metric for this black hole can be written
in the form \cite{p13,p14} 
\begin{equation}
ds^{2}=-f(r)dt^{2}+f^{-1}(r)dr^{2}+r^{2}h_{ij}dx^{i}dx^{j},
\end{equation}
with 
\begin{equation}
f(r)=\kappa+\frac{r^{2}}{2\alpha}-\frac{r^{2}}{2\alpha}\sqrt{1+\frac{64\pi\alpha M}{(d-2)\Sigma_{\kappa}r^{d-1}}-\frac{2\alpha Q^{2}}{(d-2)(d-3)r^{2d-4}}-\frac{4\alpha}{l^{2}}}\label{eq:fgaussbonnet}
\end{equation}
and 
\begin{equation}
\Lambda=-\frac{(d-1)(d-2)}{2l^{2}}.
\end{equation}
Here $\alpha=(d-3)(d-4)\alpha_{GB}$ is the redefined Gauss-Bonnet
coefficient (and $\alpha_{GB}$ the Gauss-Bonnet coefficient), $M$
and $Q$ are the mass and electric charge of the black hole, respectively,
$\Lambda$ is the cosmological constant (which we assume to be fixed),
and $r^{2}h_{ij}dx^{i}dx^{j}$ is the line element of the $(d-2)$-dimensional
maximally symmetric Einstein space with constant curvature $(d-2)(d-3)\kappa$
and volume $\Sigma_{\kappa}$. The parameter $\kappa$ defines the
topology of the black hole event horizon; for our purposes we will
focus on the case of spherical topology, corresponding to $\kappa=1$
\cite{p14,p15}. We note that because the Gauss-Bonnet term is a topological
invariant for $d<5$, we must assume $d\geq5$.

Since the metric is static and spherically symmetric, the first condition
in the theorem is satisfied, as shown in Section~\ref{sectionquintess}.
The electric potential at the event horizon $\Phi$, Hawking temperature
$T$ and entropy $S$ are given by \cite{p13} 
\begin{equation}
\Phi=\frac{Qr_{+}^{3-d}\Sigma_{\kappa}}{16\pi(d-3)},\qquad T=\frac{f'(r_{+})}{4\pi},\qquad S=r_{+}^{d-4}(2\alpha(d-2)+(d-4)r_{+}^{2})\Sigma_{\kappa}.\label{eq:entrquirgb}
\end{equation}
For fixed $\Lambda$ and $\alpha$, the quantities above satisfy the
first law of thermodynamics \cite{p13,p14} 
\begin{equation}
dM=TdS+\Phi dQ.
\end{equation}
By setting $f(r)=0$ we obtain two solutions, $r_{-}$ and $r_{+}$,
and thus the black hole has an inner and outer horizon, which coincide
as we take it to be extremal. From \eqref{eq:fgaussbonnet} we have
\begin{equation}
f(r_{+})=0\Leftrightarrow M=M(r_{+},Q),
\end{equation}
and using \eqref{eq:entrquirgb} we obtain a Smarr relation in the
form 
\begin{equation}
M=M(S,Q),
\end{equation}
which fullfills the second condition of the theorem. The third condition in
the theorem then follows by the same argument as in Section~\ref{sectionquintess}.

Finally, assuming that the test fields satisfy the null energy condition
and preserve the black hole type, we have thus shown it is not possible
to violate the weak cosmic censorship with such test fields, as all
conditions in the theorem are met.

\section{RN-AdS black hole in nonlinear electrodynamics}

For a $d$-dimensional static, spherically symmetric, electrically
charged AdS black hole solution in general nonlinear electrodynamics
(NLED) theories we have the metric and NLED field given by \cite{p16}
\begin{equation}
ds^{2}=-f(r)dt^{2}+f(r)^{-1}dr^{2}+r^{2}d\Omega_{d-2}^{2},
\end{equation}
\begin{equation}
{\bf A}=A_{t}(r)dt,
\end{equation}
where $d\Omega_{d-2}^{2}$ is the metric on the unit $(d-2)$-sphere,
${\bf A}$ is the nonlinear electromagnetic potential one-form and
$f(r)$ is given by 
\begin{equation}
f(r)=1-\frac{m}{r^{d-2}}+\frac{r^{2}}{l^{2}}-\frac{4}{d-2}\frac{1}{r^{d-3}}\int_{r}^{\infty}r^{d-2}\Big[\mathcal{L}(s;a_{i})-A'_{t}(r)\frac{q}{r^{d-2}}\Big]dr.\label{eq:fNLED}
\end{equation}
Here $m$ is a constant related to the black hole mass, $l$ is the
AdS radius, related to the cosmological constant by $\Lambda=-\frac{(d-1)(d-2)}{2l^{2}}$,
$q$ is a constant related to the black hole charge, $\mathcal{L}(s;a_{i})$
is the generic NLED Lagrangian, with $a_{i}$ characterizing the non-linearity
of the electrodynamics, and $s$ an independent nontrivial scalar.
Note that the metric approaches the usual RN-AdS metric when $r\rightarrow\infty$.

Since the metric is static and spherically symmetric, the first condition
in the theorem is satisfied, as shown in Section~\ref{sectionquintess}.
The electric potential at the event horizon $\Phi$, Hawking temperature
$T$, black hole mass $M$, entropy $S$ and black hole charge $Q$
are given by \cite{p16} 
\begin{equation}
\Phi=-A_{t}(r_{+}),\hspace{1cm}T=\frac{f'(r_{+})}{4\pi},\hspace{1cm}M=\frac{d-2}{16\pi}\omega_{d-2}m,\label{eq:entreqNLED}
\end{equation}
\begin{equation}
S=\frac{r_{+}^{d-2}\omega_{d-2}}{4},\hspace{1cm}Q=\frac{q}{4\pi}\omega_{d-2},\label{eq:entreqNELD2}
\end{equation}
where $\omega_{d-2}$ is the volume of the unit $(d-2)$-sphere. Assuming
the cosmological constant and the parameters $s,a_{i}$ from the NLED
theory to be fixed, we have the first law of black hole thermodynamics
\cite{p16} 
\begin{equation}
dM=\Phi dQ+TdS.
\end{equation}
From \eqref{eq:fNLED} together with \eqref{eq:entreqNLED} and \eqref{eq:entreqNELD2}
we see that 
\begin{equation}
f(r_{+})=0\Leftrightarrow M=M(r_{+},Q),
\end{equation}
and using \eqref{eq:entreqNELD2} we obtain a Smarr relation in the
form 
\begin{equation}
M=M(S,Q),
\end{equation}
thus satisfying the second condition of the theorem. The third condition in
the theorem then follows by the same argument as in Section~\ref{sectionquintess}.

Finally, assuming the test fields satisfy the null energy condition
and preserve the black hole type, we have shown it is not possible
to violate the weak cosmic censorship with such test fields, as all
conditions in the theorem are met.

\section{BTZ black hole in nonlinear electrodynamics }

The metric for the $(2+1)$-dimensional static BTZ black hole coupled
with nonlinear electrodynamics (NLED) is given by \cite{p17} 
\begin{equation}
ds^{2}=-f(r)dt^{2}+f(r)^{-1}dr^{2}+r^{2}d\phi^{2},
\end{equation}
where 
\begin{equation}
f(r)=\frac{r^{2}}{l^{2}}-m-q^{2}\log\Big(\frac{q^{2}}{a^{2}l^{2}}+\frac{r^{2}}{l^{2}}\Big).\label{eq:fBTZ-NLED}
\end{equation}
Here $l$ is the AdS radius, related to the cosmological constant
$\Lambda$ by 
\begin{equation}
\Lambda=-\frac{1}{l^{2}},
\end{equation}
$m$ and $q$ are parameters that relate to the black hole mass $M$
and charge $Q$ by 
\begin{equation}
M=\frac{m}{8},\qquad Q=8q,
\end{equation}
and $a$ is a free parameter. The electromagnetic potential one-form
is given by \cite{p17} 
\begin{equation}
{\bf A}=-\frac{a^{2}qr^{2}-q^{3}\log(q^{2}+a^{2}r^{2})}{32\pi}dt.
\end{equation}

Since the metric is static and spherically symmetric, the first condition
in the theorem is satisfied, as shown in Section~\ref{sectionquintess}.
The electric potential at the event horizon $\Phi$, Hawking temperature
$T$, black hole mass $M$ and entropy $S$ can be expressed as \cite{p17}
\begin{equation}
\Phi=-\frac{Q(l^{2}Q^{2}+(l^{2}Q^{2}+64a^{2}r_{+}^{2})\log(\frac{Q^{2}}{64a^{2}}+\frac{r_{+}^{2}}{l^{2}})}{256(l^{2}Q^{2}+64a^{2}r_{+}^{2})},
\end{equation}
\begin{equation}
T=\frac{r_{+}}{128l^{2}\pi}\Big(64-\frac{64a^{2}Q^{2}r_{+}^{2}}{l^{2}Q^{2}+64a^{2}r_{+}^{2}}\Big),
\end{equation}
\begin{equation}
M=\frac{r_{+}^{2}}{8l^{2}}-\frac{Q^{2}}{512}\log\Big[\frac{1}{l^{2}}\Big(\frac{Q^{2}}{64a^{2}}+r_{+}^{2}\Big)\Big],
\end{equation}
\begin{equation}
S=\frac{1}{2}\pi r_{+},\label{eq:entreqBTZ-NLED}
\end{equation}
where $r_{+}$ is the radius of the event horizon. For fixed $a$
and $\Lambda$, we have the first law of thermodynamics \cite{p17}
\begin{equation}
dM=TdS+\Phi dQ.
\end{equation}
From \eqref{eq:fBTZ-NLED} we have 
\begin{equation}
f(r_{+})=0\Leftrightarrow M=M(r_{+},Q),
\end{equation}
and using \eqref{eq:fBTZ-NLED} we obtain a Smarr relation in the
form 
\begin{equation}
M=M(S,Q),
\end{equation}
which satisfies the second condition in the theorem. The third condition in
the theorem then follows by the same argument as in Section~\ref{sectionquintess}.

Finally, assuming that the test fields satisfy the null energy condition
and preserve the black hole type, we just have shown it is not possible
to violate the weak cosmic censorship with such test fields, as all
conditions in the theorem are met.

\section{Born-Infeld-AdS black holes}

The $d$-dimensional Born-Infeld-AdS black hole solution spacetime
may be written as \cite{p18,p19} 
\begin{equation}
ds^{2}=-f(r)dt^{2}+f(r)^{-1}dr^{2}+r^{2}d\Omega_{d-2}^{2},
\end{equation}
with $f(r)$ given by 
\begin{equation}
\begin{split}f(r)= & 1-\frac{m}{r^{d-3}}+\frac{r^{2}}{l^{2}}+\frac{4b^{2}r^{2}}{(d-1)(d-2)}\left(1-\sqrt{1+\frac{(d-2)(d-3)q^{2}}{2b^{2}r^{2d-4}}}\right)\\
 & +\frac{2(d-2)q^{2}}{(d-1)r^{2d-6}}{}_{2}F_{1}\left[\frac{d-3}{2d-4},\frac{1}{2},\frac{3d-7}{2d-4},-\frac{(d-2)(d-3)q^{2}}{2b^{2}r^{2d-4}}\right],
\end{split}
\label{eq:fborninf}
\end{equation}
where $_{2}F_{1}$ is the hypergeometric function and $b$ is the
Born-Infeld parameter characterizing the nonlinearity of the electromagnetic
field. For this spacetime we have a negative constant cosmological
constant given by 
\begin{equation}
\Lambda=-\frac{(d-1)(d-2)}{2l^{2}},
\end{equation}
and the parameters $m$ and $q$ relate to the black hole mass $M$
and charge $Q$ by 
\begin{equation}
M=\frac{(d-2)\omega_{d-2}}{16\pi}m\qquad\text{ and }\qquad Q=\frac{q\omega_{d-2}}{4\pi}\sqrt{\frac{(d-2)(d-3)}{2}},
\end{equation}
with $\omega_{d-2}$ the volume of the unit $(d-2)$-dimensional sphere.
The black hole mass is given as a function of the event horizon radius
$r_{+}$ by \cite{p19} 
\begin{equation}
\begin{aligned}M= & \frac{(d-2)\omega_{d-2}}{16\pi}r_{+}^{d-3}+\frac{(d-2)\omega_{d-2}}{16\pi}r_{+}^{d-1}\\
 & +\frac{b^{2}\omega_{d-2}}{4\pi(d-1)}r_{+}^{d-1}\left(1-\sqrt{1+\frac{16\pi^{2}Q^{2}}{b^{2}\omega_{d-2}^{2}r_{+}^{2d-4}}}\right)\\
 & +\frac{4\pi(d-2)Q^{2}}{(d-1)(d-3)\omega_{d-2}r_{+}^{d-3}}{}_{2}F_{1}\left[\frac{d-3}{2d-4},\frac{1}{2},\frac{3d-7}{2d-4},-\frac{16\pi Q^{2}}{b^{2}\omega_{d-2}^{2}r_{+}^{2d-4}}\right],
\end{aligned}
\end{equation}
the Hawking temperature as 
\begin{equation}
T=\frac{1}{4\pi}\left[\frac{d-3}{r_{+}}+(d-1)r_{+}+\frac{4b^{2}r_{+}}{d-2}\left(1-\sqrt{1+\frac{16\pi^{2}Q^{2}}{b^{2}\omega_{d-2}^{2}r_{+}^{2d-4}}}\right)\right],
\end{equation}
and the entropy and electric potential at the event horizon are then
\begin{equation}
S=\frac{\omega_{d-2}r_{+}^{d-2}}{4},\label{eq:entreqborninf}
\end{equation}
\begin{equation}
\Phi=\frac{q}{\sqrt{\frac{2(d-3)}{d-2}}}\frac{1}{r_{+}^{d-3}}{}_{2}F_{1}\left[\frac{d-3}{2d-4},\frac{1}{2},\frac{3d-7}{2d-4},-\frac{(d-2)(d-3)q^{2}}{2b^{2}r_{+}^{2d-4}}\right].
\end{equation}
Since the metric is static and spherically symmetric, the first condition
in the theorem is satisfied, as shown in Section~\ref{sectionquintess}.
Assuming the Born-Infeld parameter $b$ and the cosmological constant
$\Lambda$ to be fixed, we have the first law of black hole of thermodynamics
\cite{p18} 
\begin{equation}
dM=TdS+\Phi dQ.
\end{equation}
From \eqref{eq:fborninf} we have 
\begin{equation}
f(r_{+})=0\Leftrightarrow M=M(r_{+},Q),
\end{equation}
and using \eqref{eq:entreqborninf} we obtain a Smarr relation in
the form 
\begin{equation}
M=M(S,Q),
\end{equation}
thus satisfying the second condition of the theorem. The third condition in
the theorem then follows by the same argument as in Section~\ref{sectionquintess}.

Finally, assuming that the test fields satisfy the null energy condition
and preserve the black hole type, we just have shown that test fields
cannot destroy extremal Born-Infeld-AdS black holes, as all conditions
in the theorem are met.

\section{Charged toroidal black holes}

The metric of a charged toroidal black hole is given by \cite{p20,p21}
\begin{equation}
ds^{2}=-f(r)dt^{2}+f^{-1}(r)dr^{2}+r^{2}(d\theta^{2}+d\psi^{2})
\end{equation}
with 
\begin{equation}
f(r)=-\frac{\Lambda r^{2}}{3}-\frac{2M}{\pi r}+\frac{4Q^{2}}{\pi r^{2}},\label{eq:ftorus}
\end{equation}
where the cosmological constant is negative, $\Lambda<0$. The electromagnetic
potential one-form is given by 
\begin{equation}
{\bf A}=-\frac{4Q}{r}dt.
\end{equation}
These black holes have two horizons, which coincide when the black
hole becomes extremal (the inner horizon radius $r_{-}$ becomes equal
to the outer event horizon radius $r_{+}$). The Hawking temperature
can be written as \cite{p20} 
\begin{equation}
T=\frac{f'(r_{+})}{4\pi}=\frac{-12Q^{2}+3Mr_{+}-\pi r_{+}^{4}\Lambda}{6\pi^{2}r_{+}^{3}},
\end{equation}
the black hole entropy as 
\begin{equation}
S=\pi^{2}r_{+}^{2},\label{eq:entreqtorus}
\end{equation}
and the electric potential at the event horizon is given by 
\begin{equation}
\Phi=A_{t}(\infty)-A_{t}(r_{+})=\frac{4Q}{r_{+}}.
\end{equation}

Since the metric is static, the black hole is nonrotating and there
is no angular momentum to consider; consequently, the thermodynamic
angular velocity can be set to zero. The timelike Killing vector $X=\frac{\partial}{\partial t}$
determines the black hole mass and is also the Killing generator of
the event horizon $\mathscr{H}^{+}$, whose angular velocity is then
also zero. In other words, $X=K=Z$, and so the first condition of
the theorem is satisfied.

For a fixed cosmological constant, we have the first law of black
hole thermodynamics given as \cite{p20} 
\begin{equation}
dM=TdS+\Phi dQ.
\end{equation}
The black hole mass can be obtained from \eqref{eq:ftorus} in the
form $M=M(r_{+},Q)$: 
\begin{equation}
f(r_{+})=0\Leftrightarrow M=\frac{12Q^{2}-\pi r_{+}^{4}\Lambda}{6r_{+}}.
\end{equation}
Using \eqref{eq:entreqtorus} we obtain a Smarr relation in the form
\begin{equation}
M=M(S,Q),
\end{equation}
which satisfies the second condition of the theorem. The third condition in
the theorem then follows by the same argument as in Section~\ref{sectionquintess}.

Finally, assuming that the test fields satisfy the null energy condition
and preserve the black hole type, we just have shown that it is not
possible to violate the weak cosmic censorship with such test fields,
as all conditions in the theorem are met.

\section{Charged black holes in string theory}

The metric for the RN black hole analogue solution in the low energy
limit of heterotic string theory can be written as \cite{p22,p23,p24,p25},
\begin{equation}
ds_{E}^{2}=-f(r)dt^{2}+f(r)^{-1}dr^{2}+r\left(r-\frac{Q^{2}}{M}\right)d\Omega\label{eq:fstring}
\end{equation}
where 
\begin{equation}
f(r)=1-\frac{2M}{r},
\end{equation}
$M$ is the black hole mass and $Q$ is the magnetic charge (related
to the black hole charge $q$ by $q=\frac{1}{\sqrt{4\pi Q}}$) \cite{p24}.
For this solution we have the electromagnetic potential one-form \cite{p22}
\begin{equation}
{\bf A}=-\frac{Q}{r}dt
\end{equation}
and the dilaton field $\phi$ given by 
\begin{equation}
e^{2\phi}=1-\frac{Q^{2}}{Mr}.
\end{equation}
Here we assume spacetime to be asymptotically flat, thus with a zero
cosmological constant and a dilaton field $\phi$ vanishing at infinity
\cite{p22}. In order to make sure that the action reduces to the
usual Einstein action with scalar field when the Maxwell field vanishes,
we make the scaling on the metric $g_{\mu\nu}^{E}=e^{-2\phi}g_{\mu\nu}$
\cite{p22,p23,p24,p25}.

Since the metric is static and spherically symmetric, the first condition
in the theorem is satisfied, as shown in Section~\ref{sectionquintess}.
The Hawking temperature $T$ and electric potential $\Phi$ at the
event horizon are 
\begin{equation}
T=\frac{f'(r_{+})}{4\pi}=\frac{M}{2\pi r_{+}^{2}}=\frac{1}{8\pi M},
\end{equation}
\begin{equation}
\Phi=A_{t}(\infty)-A_{t}(r_{+})=\frac{Q}{r_{+}},
\end{equation}
where in the last step to obtain the temperature we used the fact
that the event horizon radius is $r_{+}=2M$. The black hole mass
$M$ and entropy $S$ are then given by \cite{p22,p25} 
\begin{equation}
M=\frac{Q^{2}}{r_{+}},\hspace{1cm}S=\frac{A}{4}=\pi r_{+}^{2}-2\pi Q^{2}=4\pi M^{2}-2\pi Q^{2},\label{eq:entreqstring}
\end{equation}
where $A$ is the event horizon's area. These parameters are then
related through the first law of black hole thermodynamics \cite{p25}
\begin{equation}
dM=TdS+\Phi dQ
\end{equation}
(assuming a fixed cosmological constant).
From \eqref{eq:entreqstring} is clear that we have a Smarr relation
in the form 
\begin{equation}
M=M(r_{+},Q)=M(S,Q),
\end{equation}
thus satisfying the second condition of the theorem. The third condition in
the theorem then follows by the same argument as in Section~\ref{sectionquintess}.

Finally, assuming that the test fields satisfy the null energy condition
and preserve the black hole type, we just have shown that it is not
possible to violate the weak cosmic censorship with such test fields,
as all conditions in the theorem are met.

\section{5D charged rotating minimally gauged supergravity black hole}

The metric for the $5$-dimensional rotating minimally gauged supergravity
black hole solution is given by \cite{p26} 
\begin{equation}
\begin{aligned}ds^{2}= & -\left(dt-a\sin^{2}\theta d\phi-b\cos^{2}\theta d\psi\right)\left[f\left(dt-a\sin^{2}\theta d\phi-b\cos^{2}\theta d\psi\right)\right.\\
 & \left.+\frac{2q}{\Sigma}\left(b\sin^{2}\theta d\phi+a\cos^{2}\theta d\psi\right)\right]+\Sigma\left(\frac{r^{2}dr^{2}}{\Delta}+d\theta^{2}\right)\\
 & +\frac{\sin^{2}\theta}{\Sigma}\left[adt-\left(r^{2}+a^{2}\right)d\phi\right]^{2}+\frac{\cos^{2}\theta}{\Sigma}\left[bdt-\left(r^{2}+b^{2}\right)d\psi\right]^{2}\\
 & +\frac{1}{r^{2}\Sigma}\left[abdt-b\left(r^{2}+a^{2}\right)\sin^{2}\theta d\phi-a\left(r^{2}+b^{2}\right)\cos^{2}\theta d\psi\right]^{2},
\end{aligned}
\label{eq:f5D}
\end{equation}
with 
\begin{equation}
\begin{aligned}f(r,\theta) & =\frac{\left(r^{2}+a^{2}\right)\left(r^{2}+b^{2}\right)}{r^{2}\Sigma}-\frac{\mu\Sigma-q^{2}}{\Sigma^{2}},\\
\Sigma(r,\theta) & =r^{2}+a^{2}\cos^{2}\theta+b^{2}\sin^{2}\theta,\\
\Delta(r) & =\left(r^{2}+a^{2}\right)\left(r^{2}+b^{2}\right)+2abq+q^{2}-\mu r^{2}.
\end{aligned}
\end{equation}
Here $\mu$, $q$, $a$ and $b$ are the mass, charge and angular
momentum per unit mass parameters, respectively, and are related to
the physical mass $M$, charge $Q$ and angular momenta $J_{\phi}$
and $J_{\psi}$ by \cite{p26} 
\begin{equation}
\mu=\frac{8M}{3\pi},
\end{equation}
\begin{equation}
q=\frac{4Q}{\sqrt{3}\pi},
\end{equation}
\begin{equation}
a+b=\frac{4}{\pi}\frac{J_{\phi}+J_{\psi}}{\mu+q}.
\end{equation}
The electromagnetic potential one-form can be written as
\begin{equation}
{\bf A}=-\frac{\sqrt{3}q}{2\Sigma}\left(dt-a\sin^{2}\theta d\phi-b\cos^{2}\theta d\psi\right).
\end{equation}

The Killing generator of the event horizon is given by \cite{p26}
\begin{equation}
Z=\frac{\partial}{\partial t}+\Omega^{(\phi)}\frac{\partial}{\partial\phi}+\Omega^{(\psi)}\frac{\partial}{\partial\psi},
\end{equation}
with 
\begin{equation}
\begin{aligned}\Omega^{(\phi)} & =\frac{a\left(r_{+}^{2}+b^{2}\right)+bq}{\left(r_{+}^{2}+a^{2}\right)\left(r_{+}^{2}+b^{2}\right)+abq},\\
\Omega^{(\psi)} & =\frac{b\left(r_{+}^{2}+a^{2}\right)+aq}{\left(r_{+}^{2}+a^{2}\right)\left(r_{+}^{2}+b^{2}\right)+abq}.
\end{aligned}
\end{equation}
We have the first law of black hole thermodynamics \cite{p26} 
\begin{equation}
dM=\frac{\kappa}{8\pi}dA+\Omega^{(\phi)}dJ_{\phi}+\Omega^{(\psi)}dJ_{\psi}+\Phi dQ,
\end{equation}
where A is the area of the event horizon, 
\begin{equation}
A=\frac{2\pi^{2}}{r_{+}}\left(\mu r_{+}^{2}-abq-q^{2}\right),\label{eq:entreq5D}
\end{equation}
$\kappa$ is the surface gravity at the event horizon, 
\begin{equation}
\kappa=\frac{\left(2r_{+}^{2}+a^{2}+b^{2}-\mu\right)r_{+}}{\mu r_{+}^{2}-abq-q^{2}},
\end{equation}
and $\Phi$ is the electric potential at the event horizon, 
\begin{equation}
\Phi=\frac{\sqrt{3}qr_{+}^{2}}{\mu r_{+}^{2}-abq-q^{2}}.
\end{equation}
This shows that the first condition of the theorem is satisfied, since
$\Omega^{(\phi)}$ and $\Omega^{(\psi)}$ are the thermodynamic angular
velocities. From \eqref{eq:f5D} we see that 
\begin{equation}
f(r_{+})=0\Leftrightarrow M=M(r_{+},J_{\phi},J_{\psi},Q),
\end{equation}
and using \eqref{eq:entreq5D} and the Bekenstein-Hawking formula
\cite{p8}, we obtain a Smarr relation in the form 
\begin{equation}
M=M(S,J_{\phi},J_{\psi},Q),
\end{equation}
thus satisfying the second condition of the theorem.

When the black hole becomes extremal by definition we have $k=0$,
or equivalently $T=0$, leading to 
\begin{equation}
\frac{\partial M}{\partial S}(S,J_{\phi},J_{\psi},Q)=0.
\end{equation}
Solving for the entropy leads to the entropy, hence to the mass, of
an extremal black hole as a function of its angular momentum and
charge, 
\begin{equation}
S=S_{ext}(J_{\phi},J_{\psi},Q)\Rightarrow M_{ext}=M(S_{ext}(J_{\phi},J_{\psi},Q),J_{\phi},J_{\psi},Q),
\end{equation}
satisfying the third condition of the theorem.

Finally, assuming the test fields interacting with the black hole
satisfy the null energy condition and preserve the black hole type
the black hole type, all the conditions of the theorem are met and
thus the weak cosmic censorship is preserved for such test fields.

\section{(2+1)-dimensional MTZ black holes}

The metric for the (2+1)-dimensional MTZ black hole is given by \cite{p27}
\begin{equation}
ds^{2}=-f(r)dt^{2}+\frac{dr^{2}}{f(r)}+r^{2}d\phi^{2},
\end{equation}
with 
\begin{equation}
f(r)=r^{2}-M-\left(\frac{Q}{2}\right)^{2}\ln(r^{2}),\label{eq:}
\end{equation}
and the electromagnetic potential one-form given by 
\begin{equation}
{\bf A}=-Q\ln{r}.
\end{equation}

Since the metric is static and spherically symmetric, the first condition
in the theorem is satisfied, as shown in Section~\ref{sectionquintess}.
The Hawking temperature and electric potential at the event horizon
can we written as 
\begin{equation}
T=\frac{f'(r_{+})}{4\pi}=\frac{r_{+}}{2\pi}-\frac{Q^{2}}{8\pi r_{+}},
\end{equation}
\begin{equation}
\Phi=\left(\frac{\partial M}{\partial Q}\right)_{|_{r=r_{+}}}=-Q\ln{r_{+}}.
\end{equation}
We have 
\begin{equation}
f(r_{+})=0\Leftrightarrow M=r_{+}^{2}-\left(\frac{Q}{2}\right)^{2}\ln{r_{+}^{2}}=M(r_{+},Q).
\end{equation}
Noting that in $(2+1)$-dimensions the area $A$ of the black hole
horizon is the perimeter, $A=2\pi r_{+}$, and using the Bekenstein-Hawking
formula \cite{p8}, we obtain the Smarr relation in the form 
\begin{equation}
M=M(S,Q),
\end{equation}
yielding the first law of black hole thermodynamics 
\begin{equation}
dM=TdS+\Phi dQ,
\end{equation}
hence satisfying the second condition of the theorem. The third condition in
the theorem then follows by the same argument as in Section~\ref{sectionquintess}.

Finally, assuming the test fields satisfy the null energy condition
and preserve the black hole type, all the conditions of the theorem
are met and we conclude that such test fields cannot destroy extremal
(2+1)-dimensional MTZ black holes.

\section{Discussion}

We have shown, using the general result in \cite{p1},
that the weak cosmic censorship holds for various extremal black holes
in the literature; more precisely, it is impossible to create naked singularities
by overcharging or overspinning these extremal black holes, as long as the test
particles or fields interacting with them satisfy the null energy condition.
Although we chose a small subset of the many extremal black holes
in the literature, the theorem in \cite{p1} can be used to check
the validity of the conjecture for other extremal black holes as well 
(recent interesting examples include \cite{Gwak, YCWL}).

Our results seem to indicate that as long as the interacting test
particles or fields satisfy the null energy condition and only exchange
energy, angular momentum and electric charge with the black hole 
then the conclusion that weak cosmic censorship holds
is almost automatic  (assuming that there is a first law of 
thermodynamics relating the changes in these 
quantities). In fact, the best strategy to look for possible
violations of weak cosmic censorship appears to be either to use test
fields which do not satisfy the null energy condition (such as fermion
fields \cite{Duztas,Toth}) or to use test fields that exchange other
types of conserved charges with the black hole.

We should emphasize that we assumed that the interaction
with the test fields preserves the black hole type, meaning that after the 
interaction the spacetime settles to a stationary black hole solution 
characterized by the same parameters. In this sense our proof is not exhaustive,
since there is always the possibility of violating the weak cosmic censorship conjecture
via interactions that change the black hole type.

\section*{Acknowledgements}
This work was partially supported by FCT/Portugal through projects 
UID/MAT/04459/2019 and UIDB/MAT/04459/2020
and grant (GPSEinstein) PTDC/MAT-ANA/1275/2014.

\end{document}